**TITLE: THE INSEPARABILITY OF SAMPLING AND TIME AND ITS INFLUENCE ON ATTEMPTS TO UNIFY THE MOLECULAR AND FOSSIL RECORDS**

**By:**
Melanie J. Hopkins[1], David W. Bapst[2], Carl Simpson[3], and Rachel C.M. Warnock[4]

[1]Division of Paleontology, American Museum of Natural History, Central Park West at 79th Street, New York, NY, 10024, USA. mhopkins@amnh.org
[2]Department of Ecology and Evolutionary Biology, University of Tennessee, Knoxville, TN 37996-1610, USA & Department of Geology and Geophysics, 3115 TAMU, Texas A&M University, College Station, Texas 77843, USA. dwbapst@tamu.edu
[3]University of Colorado Museum of Natural History and Department of Geological Sciences, University of Colorado, Boulder 265 UCB Boulder, CO 80309 USA. Carl.Simpson@colorado.edu
[4]Department of Biosystems Science and Engineering, ETH Zurich, Mattenstraße 26, 4058 Basel, Switzerland. rachel.warnock@bsse.ethz.ch

**Abstract:**

The two major approaches to studying macroevolution in deep time are the fossil record and reconstructed relationships among extant taxa from molecular data. Results based on one approach sometimes conflict with those based on the other, with inconsistencies often attributed to inherent flaws of one (or the other) data source. What is unquestionable is that both the molecular and fossil records are limited reflections of the same evolutionary history, and any contradiction between them represents a failure of our existing models to explain the patterns we observe. Fortunately, the different limitations of each record provide an opportunity to test or calibrate the other, and new methodological developments leverage both records simultaneously. However, we must reckon with the distinct relationships between sampling and time in the fossil record and molecular phylogenies. These differences impact our recognition of baselines, and the analytical incorporation of age estimate uncertainty. These differences in perspective also influence how different practitioners view the past and evolutionary time itself, bearing important implications for the generality of methodological advancements, and differences in the philosophical approach to macroevolutionary theory across fields.

"Incompleteness and informativeness are not strictly coupled." Donoghue et al. (1989)

"Time is on my side, yes it is. / Time is on my side, yes it is." Jerry Ragovoy (1963)

Both paleontologists and comparative biologists would like to be able to look into the past to understand the history of life on Earth and the underlying evolutionary processes that led to the biological world of today. Both disciplines collect evidence about the nature of ancient life in deep time. For paleontologists, the evidence comes primarily from the fossil record; for comparative biologists, the evidence comes primarily from living organisms. The inferences drawn from one record sometimes conflict with the other, but today many paleontologists and comparative biologists are working to reconcile these apparent discrepancies. One way of doing this is to explicitly incorporate fossil taxa into molecular phylogenetic hypotheses and ensuing comparative analyses. While emerging 'tip-dating' methods promise to change the face of systematic biology and paleontology (Gavryushkina et al., 2016; Donoghue and Yang 2016; Hunt and Slater 2016; Wright 2017), we need to directly address how the relationship between sampling and time differ between the fossil record and the 'molecular record', by which we mean the record of evolutionary processes found through the phylogenetic analysis of molecular data. Comparative biologists and paleontologists use different types of data that vary fundamentally in their association with time, and this discrepancy has far-reaching and under-appreciated implications for macroevolutionary analysis. Beyond impacting how we derive accurate ages for individual specimens or calibrate divergence dates, this discrepancy complicates efforts to reconcile, compare, or combine these two records, and influences how comparative biologists and paleontologists view the past. In this paper we summarize some recent work that demonstrates how these two records can complement each other in spite of their respective limitations. We then outline challenges that remain, focusing particularly on those that arise in the context of sampling and time, and propose some research directions for overcoming them.

**Two limited but complementary records of life**

Paleontologists directly observe the remains and traces of life contained within rocks formed long ago. However, paleontologists must work with an incomplete picture: most individual organisms are not preserved or sampled, and those that have been are represented by only some anatomical parts and are usually chemically altered. In contrast, comparative biologists can employ any number of ecological, molecular, morphological, developmental and behavioral approaches to study change in modern species and populations. As difficult as it may be to collect samples of living organisms, it is possible to do so comprehensively. However, as comparative biologists move their perspective backwards in time, their inferences become more dependent on extrapolating observations from living organisms, across reconstructed relationships from molecular data, and thus their view of the past moves out of focus and loses precision as they look further back in time. Both records have limitations that will never be completely surpassed.

However, what is missing from one record is often available from the other. The fossil record supplies a direct record of past diversity that frequently includes character combinations, ecological associations, and distributions that are not inferable from analyses of recent taxa alone. For example, recent molecular phylogenies of mammals (e.g., dos Reis et al. 2012) predict that the two families of modern sloths diverged between 50 and 20 Ma. Morphological phylogenetic analysis including fossils makes the same prediction (Raj Pant et al. 2014). In addition, the ancestral body size that would be reconstructed from modern species is similar to that of the oldest fossil of this group (*Pseudoglyptodon*, see Raj Pant et al. 2014, fig. 2). Thus in terms of the age and the body size of the ancestral root, the molecular and fossil records are incredibly consistent. However, between 30 Ma and the present, the fossil sloth species (of which there are 51) included a large range in body size, some three orders of magnitude greater in body size than any extant species (of which there are 6) (Raj Pant et al. 2014). Similarly, dated molecular phylogenies suggested that lignin-decomposing fungi did not evolve until the Permian (Floudas et al. 2012; Kohler et al. 2015). This implies that that the abundant coal deposition of the Carboniferous could have been caused by the absence of these lignin decomposers. Nelsen et al. (2016) refute this by documenting the presence of lignin-decomposition in the fossil record as early as the

Devonian, and by noting that the primary source of Carboniferous coal is the lignin-poor periderm in fossil lycopsids.

Our understanding of highly diverse modern clades is also dramatically improved by fossil data, particularly for those with long evolutionary histories. For example, arthropods are tremendously diverse today (Minelli et al. 2013), but much of known fossil arthropod diversity--both taxonomic and morphological--exists in completely extinct clades (Edgecombe and Legg 2013; Smith and Marcot 2015). In fact, morphological diversity was as great in the Cambrian (~510 million years ago) as it is among modern arthropods (Briggs et al. 1992). The morphological diversity of the Cambrian was dominated by groups that are long extinct, but the fossil record also provides insight on living arthropod clades: for example, crustacean morphological diversity increased substantially from the Cambrian to the early Carboniferous (~350 million years ago) but has remained relatively constant since that time (Wills 1998). Molecular phylogenetics has made considerable progress in deciphering relationships within and among major arthropod clades, but it is becoming increasingly evident that obtaining a more accurate arthropod tree—and a deeper understanding of trait homology and evolution, divergence times, and clade origin and radiation—will require incorporating fossil taxa (Edgecombe 2010; Legg et al. 2013; Garwood et al. 2014; Yang et al. 2016). This is surely true for other groups as well, particularly those with deep divergences among extant clades or considerable past diversity (e.g., Donoghue et al. 1989; Littlewood and Smith 1995; Springer et al. 2001; Crane et al. 2004; Hermsen and Hendricks 2008; Gauthier et al. 2012; Slater et al. 2012; Gavryushkina et al. 2015; Sutton et al. 2015). A similarly diverse set of extinct cnidarians must also have existed that remain invisible in the fossil record due to lack of skeletons, and invisible to the molecular phylogenetic record due to lack of living representatives (Tweedt and Erwin 2015).

Complementary to this, through comparative analysis, the molecular record can offer insight on what taxa might be missing from the fossil record, especially for certain time periods and environments. This is obviously the case for groups that have low preservation potential, but is also true for groups that are well represented in the fossil record. For example, molecular phylogenies suggest that photosymbiosis in corals has been repeatedly gained and lost over their evolutionary history (Barbeitos et al. 2010; Kitahara et al. 2010). This pattern of gain and loss means that the commonly deep-sea azooxanthellate corals must have had a deep and diverse evolutionary history, and that zooxanthellate and azooxanthelate corals have been nearly equally diverse over the last 200 million years (Simpson 2013). The fossil record has not historically supported this (Stanley and Cairns 1988; Gill et al. 2004), possibly because of lower preservation of deep-water sediments compared to shallow water sediments in the rock record. However, more recent work has closed the gap in diversity estimates of these two coral groups as the deep-sea record that does exist becomes more evenly sampled (Kiessling and Kocsis 2015).

Not only do the fossil and molecular records differ from one another in their limitations, these limitations are independent of one another. This is not always readily apparent: often there is similarity in the identity of the taxa that can be sampled from each record, implying similar systematic biases. This is, however, largely coincidental. For example, it may seem that charismatic species in easy-to-reach places that are abundant in the present day will likely have very good molecular and fossil records, as happens to be the case for canids (e.g., Slater 2015). However, taxa that are abundant now were not necessarily abundant in the past, nor are their fossils necessarily preserved in easy-to-reach places or in the same places that their living descendants occupy (e.g., Mayr 2004). That both living and fossil species are charismatic may drive funding opportunities, and thus increase sampling effort of both records, but this may not be enough to overcome other differences. The same can be said for other putative examples that might influence sampling and thus perceived diversity, such as geographic breadth or a tendency for one clade to be taxonomically oversplit.

This independence can be used to advantage. In the case of reef-dwelling scleractinians, there is taxonomic uncertainty and preservational/sampling issues in the fossil record, and poor taxon sampling and uncertainty in topology and node-age reconstructions based on the molecular record. Given this and difficulty in estimating extinction from molecular phylogenies, Simpson et al. (2011) expected there to be differences between molecular- and fossil-derived rates of extinction. So they focused instead on the temporal pattern of net diversification rates (i.e., the speciation rate minus extinction rate) in both the species-level fossil occurrence data and larger molecular datasets, effectively comparing the molecular-clock-derived node ages against the fossil record. This approach revealed that the molecular and fossil records of reef corals show strikingly similar patterns of diversification, and that despite the limitations of either record, neither is flawed to the point of distortion.

**Current challenges: time vs sampling**

*Time and sampling in the fossil record*

Both paleontologists and comparative biologists tend to think about each individual fossil specimen as having an age, usually in millions of years ago—and they do, but it is rarely possible to determine the true, precise age. This is not only due to 'measurement error', such as that represented by confidence intervals around radiometric dates; rather the imprecision of geologic dating is a consequence of sedimentary processes and the limitations of radiometric dating (see Patzkowsky and Holland 2012). Radiometric dates can be obtained directly from fossil material, but only if specimens are of a certain type and not very old (less than 50,000 years for carbon dating, and 500,000 years for uranium-thorium dating). Rock units, such as volcanic tuffs, that can provide much older radiometric ages for calibrating geologic time intervals are relatively rare, non-uniformly deposited in space and time, and typically non-fossil-bearing. The sedimentary rocks that preserve fossils are lithified piles of sand and mud that accumulated over time. Thus, to estimate an age for most fossils, we have to first determine the relative order of deposition of those sediment deposits. Even though the relative age of fossils collected from successive beds at a single outcrop is obvious due to their order from oldest at the bottom to youngest at the top, the relative age of fossils sampled from different localities can be difficult to determine and depend primarily on some similarity in the fossil species preserved. Because rates of sedimentation vary and sediment deposition is locally sporadic, sedimentary beds of the same thickness may represent different amounts of time, and the boundaries between beds may represent different amounts of unpreserved time, even at the same outcrop. The global geologic record is compiled by matching beds or boundaries between beds over large spatial areas, using the fossils contained within them ('biostratigraphy'), stable isotope data ('chemostratigraphy'), and regional sedimentary features related to sea level changes ('sequence stratigraphy'). An absolute age model is applied primarily by assigning radiometric dates to boundaries that represent the relative position in the sedimentary sequence of dated volcanic beds, where they can be located. A minority of stage boundaries have direct dates, which means that most boundary ages are estimated by interpolation (Gradstein et al. 2012). The preservation of magnetic reversals in association with seafloor spreading has also been used for the last 50 years to calibrate the geologic time scale but can only be applied from the Late Jurassic to the present. Orbital tuning is selectively replacing such 'magnetochron' methods, but has its greatest application to deep sea marine sediments from the Oligocene to the present (Gradstein et al. 2012). Despite these challenges, it is possible to build high-resolution global timelines of fossil occurrences scaled to absolute time, including into the Paleozoic (e.g. Sadler 2004; Sadler et al. 2009), but the resolution of the geological timescale—and the number of fossil occurrences that can be placed within it—varies from study to study, and high-resolution timelines exist mainly for datasets limited to specific taxonomic groups.

So for most fossil specimens, what can be determined is the particular interval of geologic time from which that specimen was collected, but not its exact age. The length of this interval can vary considerably. Highly precise geologic records can have very short intervals encompassing single years, in the case of some lake deposits; to thousands of years for deep-sea sediment cores. Conversely, some records can have intervals encompassing staggeringly long amounts of time, such as tens of millions of years for some terrestrial deposits. Across the Phanerozoic, the global marine time scale has an average resolution of 5.5 million years; for just the Cenozoic, this improves to an average of 3.7 million years.

Fossil taxa are often represented by multiple specimens sampled from multiple collections spanning multiple stratigraphic intervals, giving that taxon a stratigraphic 'range' which represents its temporal duration. The oldest and youngest specimens that have been sampled for a fossil taxon represent the 'first occurrence' or 'first appearance', and the 'last occurrence/appearance', respectively, of that taxon. The same age uncertainty that exists for any fossil specimen (see above) also applies to these specimens; specifically, the age of the first and last appearances can be bracketed within particular intervals of time but precise ages can only rarely be assigned. In contrast, how well the sampled first and last appearances represent the 'true' first and last appearances of the taxon is also an issue of accuracy and has been the subject of considerable paleobiological research (see *Stratigraphic ranges and stasis* below).

*Time and sampling in the molecular record*

In comparison to the fossil record, the relationship between sampling and time is straightforward for living taxa: almost all individuals sampled have a timestamp of sometime in the past 150 years, with very rare exceptions (Leonardi et al. 2017). At the scale of the evolutionary histories of clades, these individuals are effectively the same age, and we denote this by saying that they were sampled from the Recent.

*Assigning times of observation*

Sampling-time relationships influence how dates are assigned to different biological units. In dated molecular phylogenies, branch tips are inherently linked to observational data sampled from contemporaneous populations. Many comparative approaches assume that the age of tip taxa are single point estimates, even if non-contemporaneous data is allowed. This concept of branch tips as instantaneous populations is incongruous with the paleontological concept of species, because morphologically-indistinct specimens sampled across millions of years of geologic time are often treated as a single contiguous taxonomic unit (e.g., a 'morphospecies') with a stratigraphic range. Paleontological data thus complicates any analysis where a taxon needs to be assigned a single precise age, requiring some treatment to deal with persistent morphotaxa, and leading to the so-called 'times of observation' problem (Fig. 1; Bapst 2013). Previous studies have assigned fixed ages by sampling from a uniform distribution bounded by the taxon stratigraphic range (e.g., Heath et al. 2014), but the stratigraphic range is distinct from stratigraphic uncertainty in the age of a specimen or the estimated first occurrence of the taxon (see *Time and sampling in the fossil record* above), and the two should not be confounded.

The estimated time of observation that should be used in a given analysis will depend on the parameters of interest. For example, if one is interested in the morphological evolution of a continuous trait, the appropriate time of observation may be the estimated age of the particular specimen from which the morphological data was measured. But if one is interested in discrete character change along branches, then the appropriate time of observation may be the first known appearance of the taxon, since all of the discrete characteristics that describe that taxon must be present by that point (e.g., Hopkins and Smith 2015). For living taxa with a fossil record, this complicates efforts to combine morphological and molecular data in single analyses, as the morphological data might be more appropriately tied to the first fossil appearance of a species while the molecular data is tied to the present day. The confounding effect of different times of observation is largely unexplored in the literature, but choices about times of observation should not be ignored, as these decisions can impact phylogenetic comparative analyses with fossil data (Bapst 2014; Cuff et al. 2015). A similar issue arises with calculations of phylogenetic diversity and the loss of evolutionary history, the measurement of which make implicit assumptions about how we conceptualize the persistence of taxonomic units over geologic time (Huang et al. 2015).

*Stratigraphic ranges and stasis*

Paleontologists have long worried about how well the age of the oldest (or youngest) sampled specimens of a species represent the 'true' first (or last) appearance of that species (Strauss and Sadler 1989; Marshall 1990, 1997; Weiss and Marshall 1999; Holland 2003, to name a few; see Wang and Marshall 2016 for a review). The gap between speciation and first observation of a species is due at least in part to the evolutionary history of a species or clade, rather than simply a general failing of the fossil record. This is because most species appear to first originate and go extinct with small geographic ranges and low abundance (Foote et al. 2007; Liow and Stenseth 2007; Foote et al. 2008), which may reduce the probability that member individuals will be sampled, even if they are preserved somewhere in the rock record (e.g. CoBabe and Allmon 1994). High-resolution sampling of the fossil record can only occasionally provide information about the process of speciation, or at least about the pattern of morphological divergence among putative descendent populations from an ancestral stock (e.g. Geary 1992; Hunt et al. 2008).

Paleontologists tend not to consider where in the process of reproductive isolation that morphological distinctiveness occurs, and thus what stage of speciation is reflected by such first appearances, as the geologic record often involves coarse timescales. However, there is considerable variation in how quickly speciation occurs (regardless of how it is defined, Marie Curie Speciation Network 2012), with some estimates as long as millions of years (to complete reproductive isolation, e.g. Dufresnes et al. 2015). The appearance across clades of morphologically distinct units, as recognized in paleontology, may be entirely uncoupled from when lineages diverge (e.g., relative to divergences reconstructed on a dated molecular phylogeny, Huang et al. 2015). The question of whether most morphological differentiation occurs during speciation is still under active investigation (Pennell et al. 2014) but has implications for estimating evolutionary rates and divergence times, since we lack phylogenetic models that incorporate this possibility.

There will always be an interval of time over which that taxon existed, even if the first and last appearances of a taxon are known to approximate the time of speciation and extinction both accurately and precisely. This is the basic observation underlying the concept of stasis. Over many years, there has been discussion about how stasis should be modeled, how sampling and species concepts influence the documented patterns, and what processes might be driving it (see Lidgard and Hopkins 2015, for an annotated bibliography). Such considerations influence how stasis has been defined, with definitions ranging from the more general and oft-used pattern-based 'little net morphological (or evolutionary) change within a lineage' (Eldredge et al. 2005; Pennell et al. 2014), to the highly specific, process-based "no evolution within any of the coexisting species due to their interactions with their biotic or abiotic environment, but with occasional minimal evolution due to genetic drift" (Nordbotten and Stenseth 2016). Despite this debate on quantifying stasis, the foundational observation for the concept of stasis—that 'morphospecies' are consistently found within sedimentary rocks spanning multiple intervals, and thus appear to persist for sometimes millions of years—has been widely documented (see Erwin and Anstey

1995; Hallam 1998; Jackson and Cheetham 1999; Jablonski 2000; Eldredge et al. 2005; Hunt 2007; Hopkins and Lidgard 2012; Hunt et al. 2015) indicating a need to consider this aspect of the evolutionary process in phylogenetic analyses incorporating fossils.

*Shifting vantage points*

In some ways, molecular data might seem much more powerful than fossil data in that they can be treated as precisely-known points of observation. Unfortunately, the simple relationship between sampling of Recent specimens and time does not provide a universal framework for comparative analyses based on molecular phylogenies. A relatively well-known example of this comes from recent exploration of the performance of the gamma statistic. The gamma statistic assesses how internode distances vary through time relative to what would be expected under a pure birth process (Pybus and Harvey 2000). Significantly negative gamma values indicate that nodes are concentrated near the root of the tree, which in turn indicates that the diversification rate has decreased over time, as might be expected if diversification is density-dependent. However, phylogenies driven by diversity-dependent diversification will only yield significantly negative gamma values if the tree is sampled around the time that equilibrium diversities were first reached (Liow et al. 2010), and thus carries poor test power when the present marks a different point in a clade's diversification history.

So, although typically molecular phylogenies consist of taxa sampled from the same time, that particular time—the Recent—likely represents different points in the diversification history of different clades, contingent on both the age of the group and its specific diversification trajectory. This phenomenon is similar to the 'shifting baselines' phenomenon recognized in ecology and conservation biology. The 'shifting baselines' concept originated in fisheries research to describe the tendency for each new generation to consider the depauperate fish communities they grew up with as pristine (Pauly 1995). Thus the 'baseline' for judging the health of current populations is frequently based on personal experience, as if no relevant change had occurred before, and thus shifts generation-by-generation. In the case of molecular phylogenies, the present represents many different baselines, depending on the clade of interest. If all diversification histories could be somehow standardized to the equivalent starting point, it would become clear how the present actually represents snapshots of many different evolutionary points. This is effectively what is done to apply the gamma statistic, but ironically, the suite of snapshots represented by the present makes clade histories much harder to compare within the same framework, as is apparent from the shortcomings of the gamma statistic. With the shifting baselines concept in mind, we have dubbed this the 'shifting vantage points' phenomenon.

The 'shifting vantage points' phenomenon extends to other common analyses in evolutionary biology, notably in the application of early burst (EB) models of trait evolution to phylogenetic datasets, which tests if trait evolution was higher early in a clade's history rather than later (Harmon et al. 2010; Slater and Pennell 2014). Like gamma, EB is a time-dependent model with constant change, and thus the initiation point is not arbitrary. Unfortunately, application of EB often assesses the entire dataset (i.e., all taxa on a given phylogeny) rather than a specific point chosen a priori. This means the initiation of the early burst is often at the divergence between a major clade and a small outgroup, which may have little relationship to where some prior expectation of an EB may have been placed. For molecular phylogenies, this often means the initiation of a crown clade, even though the onset of an increase in the rate of trait change and the subsequent slowdown might be earlier, in an unseen stem portion of the clade (Slater and Pennell 2014). Even for paleontological phylogenies, the tested initiation point is likely at the origin of some major named group, which may be a poor choice if clades originate long before they enter a phase of intense diversification and trait change (e.g. Cooper and Fortey 1998; Hopkins and Smith 2015). Fortunately, very recent work extends these models to allow for early bursts within subclades rather than across the entire tree (Puttnick 2018).

**Where to go from here?**

If fossils and living species are treated as observations of the same underlying diversification process, incomplete sampling can be modeled explicitly in the estimation of phylogenetic parameters (e.g., 'the fossilized birth-death process', Stadler 2010; Heath et al. 2014). Under this modelling framework, discontinuous fossil sampling, and the instantaneous sampling of extant species at the present, are treated as distinct processes. In addition, the assignment of one or more specimens to a single taxon (i.e. sampled stratigraphic ranges) can also be incorporated into extensions of this model (Stadler et al. 2017). If character data is available for both living and fossil species, the phylogenetic position of the fossils can also be inferred, including the identification of sampled ancestors (Gavryushkina et al. 2015; Zhang et al. 2016). Furthermore, the application of hierarchical models in phylogenetics enables the uncertainty associated with specimen ages to be incorporated into analyses explicitly (Drummond and Stadler 2016).

Many advances in tree building, however, have been developed mostly with molecular data in mind, which means models of morphological evolution—and our understanding of the behavior of those models—are often lacking in Bayesian and maximum-likelihood phylogenetic software. Thus the inclusion of fossil taxa in phylogenetic analysis requires improved models of morphological character evolution (Giribet 2015; Lee and Palci 2015; Sansom 2015; Wright et al. 2016). For example, existing models cannot effectively accommodate stasis or account for the possibility that morphological divergence occurs at the time of speciation. Model development should also account for the non-random nature of character preservation and sampling in the fossil record (Sansom et al. 2010; Sansom and Wills 2013; Murdock et al. 2016). This aspect of phylogenetic modeling should therefore benefit from a better appreciation of the evolution and development of morphological characters (Chipman 2015), the decay process (Sansom et al. 2013; Murdock et al. 2014), and the way in which fossil and morphological data is collected (Mounce et al. 2016; and see below). Similarly, many phylogenetic comparative methods have been developed to handle situations where the framework-tree is ultrametric (i.e. all tips are sampled from the same time point). Many available methods are clearly limited to ultrametric trees, particularly those that involve the classic reconstructed birth-death process model (e.g., the large body of BiSSE-type approaches developed to separate trait-dependent diversification from asymmetric trait change, Maddison et al. 2007; FitzJohn 2012). Unfortunately, such limitations are not always explicitly stated or appreciated. In fact, some methods have been designed such that non-ultrametric trees are inappropriate, even though the analysis itself is applicable in theory (see example and solution in Slater 2013, 2014). It is important for authors to clearly note when methods are not applicable to non-ultrametric phylogenies, and explain what obstacles exist to generalizing the method (as recently exemplified by Bastide et al. 2018).

Fossil occurrence data can contribute to phylogenetic parameter estimates even if character data is unavailable (Gavryushkina et al. 2014; Heath et al. 2014), meaning extinct diversity can be used to inform other macroevolutionary parameters, including speciation, extinction and species sampling. In this respect, recent developments in phylogenetics are similar to previous paleontological approaches used to generate probabilistic estimates of divergence times based on speciation, extinction and species sampling rates (Foote et al. 1999; Wilkinson et al. 2011; Nowak et al. 2013; Bapst, 2013), except that they have the advantage of being able to inform diversification rates and/or divergence times across the entire tree topology and multiple time intervals simultaneously (Stadler 2010, Gavryushkina et al. 2014). Rates of sampling and diversification during different intervals also have the potential to be better informed by available paleontological or geological data, such as regional or eustatic sea level changes, or alternative diversity and sampling proxies (Holland 1995, 2000; Smith 2001; Smith and McGowan 2007; Wagner and Marcot 2013). Indeed, fossil occurrences are the raw data paleontologists have been using for decades to infer evolutionary rates and to test virtually all models of Phanerozoic diversification (e.g., Raup 1972; Sepkoski 1981; Peters and Foote 2001; Smith 2001; Smith and McGowan 2007; Alroy et al. 2008)!

It will always be difficult to sample the fossil record uniformly in time, locally (at a single outcrop) or globally (across dated fossil occurrences), because a majority of the geologic time scale consists of intervals with relatively precise boundary dates but variable in their duration. The non-uniform nature of the fossil record violates assumptions of many phylogenetic approaches and makes it difficult to infer times of origination and extinction. The severity of such violations may be a function of taxonomic or temporal scale, but this is unexplored. In the meantime, continued increase in our understanding of sedimentary processes and the stratigraphic record make it possible to construct more realistic models for predicting local and regional fossil occurrences (e.g., Holland 1995, 2000, 2003, 2016; Hannisdal 2006), for estimating preservation and sampling rates (Foote 1997; Wagner and Margot 2013); and for evaluating hypotheses given the incompleteness of the fossil record, such as confidence that a set of taxa truly went extinct before or at some extinction event (e.g., Strauss and Sadler 1989; Marshall 1997; Hayek and Bura 2001; Wang et al 2016).

Finally, the modelling of other aspects of species evolution linked to phylogenetic history is increasingly enabled by the introduction of other process-based models in Bayesian phylogenetics, such as those for modelling biogeography (Landis 2016) or trait diversification (Kostikova et al. 2016). These developments mirror recent advances in paleontological modelling for estimating macroevolutionary parameters from incomplete and non-uniformly sampled fossil data (Silvestro et al. 2014; Brocklehurst 2015; Dunhill et al. 2016; Silvestro et al. 2016; Starrfelt and Liow 2016; Wang et al. 2016). This demonstrates a shift in both disciplines towards a more mechanistic approach to addressing questions in deep time that incorporates parameters that reflect our understanding of these evolutionary processes.

**Summary**

The data used by comparative biologists and paleontologists vary in their association with time, and in their inherent limitations. As a consequence, scientists within each field of study have developed different perspectives on how to view the past, leading to conflicting answers for some questions. The discrepancies are often tempting to reconcile by dismissing one set of observations as being wrong, rather than the more difficult solution of considering all available theory and data. Although neither the molecular record nor the fossil record are perfect, the two records bear independent limitations, and what is missing from one is often available from the other. We must deal explicitly with the different and sometimes complex relationships between time and sampling to take full advantage of the complimentary nature of the two records. Nevertheless, it is worth it: all evolutionary processes occur over time; time is what links all limited records of those processes.

**Acknowledgements**

This paper was inspired by the many illuminating discussions each of us has had with colleagues at Evolution, GSA, SVP, and other conferences. Thank you also to D. Rabosky, M. Pennell, and others, whose detailed and thoughtful criticisms of the manuscript greatly helped us crystalize our perspectives on this issue.

22Hopkins, M. J. and A. B. Smith. 2015. Dynamic evolutionary change in post-Paleozoic echinoids, and the importance of scale when interpreting rates of evolution. Proc Natl Acad Sci USA 112:3758-3763.

Huang, D., E. E. Goldberg, and K. Roy. 2015. Fossils, phylogenies, and the challenge of preserving evolutionary history in the face of anthropogenic extinctions. Proc Natl Acad Sci USA 112:4909-4914.

Hunt, G. 2007. The relative importance of directional change, random walks, and stasis in the evolution of fossil lineages. Proc Natl Acad Sci USA 104:18404-18408.

Hunt, G., M. A. Bell, and M. P. Travis. 2008. Evolution toward a new adaptive optimum: phenotypic evolution in a fossil stickleback lineage. Evolution 62:700-710.

Hunt, G., M. J. Hopkins, and S. Lidgard. 2015. Simple versus complex models of trait evolution, and stasis as a response to environmental change. Proc Natl Acad Sci USA 112:4885-4890.

Hunt, G. and G. J. Slater. 2016. Integrating paleontological and phylogenetic approaches to macroevolution. Annu Rev Ecol Evol Syst 47:189-213.

Jablonski, D. 2000. Micro- and macroevolution: scale and hierarchy in evolutionary biology and paleobiology. Pp. 15-54 *in* D. H. Erwin, and S. L. Wing, eds. Deep Time: Paleobiology's Perspective. Paleobiology 26 (suppl).

Jackson, J. B. C. and A. H. Cheetham. 1999. Tempo and mode of speciation in the sea. Trends Ecol Evol 14:72-77.

Kiessling, W. and Á. T. Kocsis. 2015. Biodiversity dynamics and environmental occupancy of fossil azooxanthellate and zooxanthellate scleractinian corals. Paleobiology 41:402-414.

Kitahara, M. V., S. D. Cairns, J. Stolarksi, D. Blair, and D. J. Miller. 2010. A comprehensive phylogenetic analysis of the Scleractinia (Cnidaria, Anthozoa) based on mitochondrial CO1 sequence data. PLoS ONE 5:e11490.

Knowlton, N. 1986. Cryptic and sibling species among the decapod Crustacea. J Crustacean Biol 6:356-363.

**Figure Captions**

**Figure 1.** Schematic showing potential variation in 'points of observation' for different types of data. Solid black line = known temporal duration based on the fossil record. Dashed black line = range extension of living taxa into the future (it is currently unknown when the last appearance will be). Blue lines = points of observation for discrete character data describing a species. Because all individuals of the species share these characters, they must be expressed by the time of the first appearance of the species. Red lines = points of observation for molecular data; tied to the age of the sampled specimen, almost always at the present except for rare samples from ancient DNA. Star = points of observation for continuous morphological data; tied to the age of the measured specimen. There may be more than one such specimen within the stratigraphic range of the species, and the age of the measured specimen(s) may not be coincident with the first or last appearance of the species.

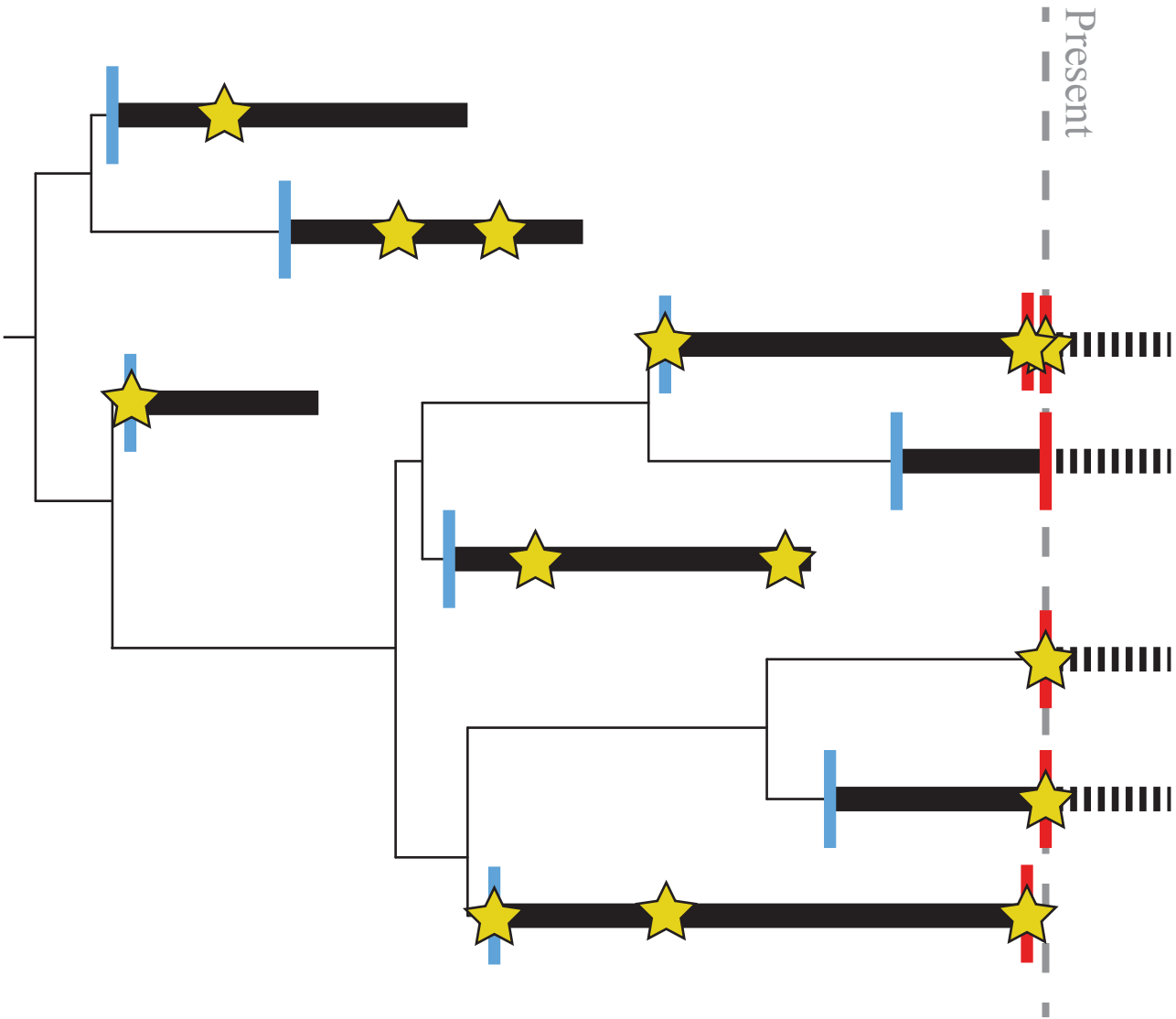